\documentclass[11pt,twocolumn,english]{article}
\usepackage[T1]{fontenc}
\usepackage[latin1]{inputenc}
\usepackage{geometry}
\usepackage{graphicx}

\makeatletter

\providecommand{\LyX}{L\kern-.1667em\lower.25em\hbox{Y}\kern-.125emX\@}

\usepackage{babel}
\makeatother
\begin{document}

\title{Narses: A Scalable Flow-Based Network Simulator}

\author{TJ Giuli and Mary Baker\\
\{giuli, mgbaker\}@cs.stanford.edu}

\maketitle
\begin{abstract}
Most popular, modern network simulators, such as ns, are targeted
towards simulating low-level protocol details. These existing simulators
are not intended for simulating large distributed applications with
many hosts and many concurrent connections over long periods of simulated
time. We introduce a new simulator, Narses, targeted towards large
distributed applications. The goal of Narses is to simulate and validate
large applications efficiently using network models of varying levels
of detail. We introduce several simplifying assumptions that allow
our simulator to scale to the needs of large distributed applications
while maintaining a reasonable degree of accuracy. Initial results
show up to a 45 times speedup while consuming 28\% of the memory used
by ns. Narses maintains a reasonable degree of accuracy -- within
8\% on average.
\end{abstract}

\section{Introduction}

With the recent surge of interest in peer-to-peer computing, the need
to simulate very large scale distributed applications over Internet-like
topologies has become apparent. Simulation results of applications
such as Gnutella and Freenet exist \cite{hong2001,zhan2002}, but
these simulations do not model network behavior such as propagation
delay, traffic interdependancies, and congestion. An example of a
peer-to-peer application that may have benefitted from network simulation
is Gnutella. In August of 2000, the Gnutella network collapsed and
fragmented into smaller networks because the flood-based nature of
its file queries overwhelmed peers with lower bandwidths\cite{clip2}.
As a result, the total amount of data available to any peer in the
network decreased drastically. If Gnutella had been simulated with
a large number of nodes, this problem could have been addressed in
the design phases of the protocol rather than \textit{ex post facto}.

Currently, network simulators are used mainly to study lower-layer
network dynamics. These simulators can characterize the behavior of
different flavors of TCP or different IP routing protocols. Because
they are targeted towards the study of lower layers, they have very
detailed models of the lower network stack. However, because of these
simulators' level of detail at lower layers, simulating applications
like Gnutella or higher-layer protocols is very expensive.

Network simulators such as ns \cite{ns} are generally discrete-event
packet-based simulators. This means that every packet in the simulation
becomes an event that must be managed in a central event queue. As
described by Ahn and Danzig \cite{Ahn1996}, the size of the event
queue in discrete event packet simulators can have a serious impact
on the performance of the simulator.

Furthermore, each packet must be processed by detailed models of network
layers, which consumes processing time. Researchers wanting to know
the behavior of a simulated application may not care to know detailed
information about the lower layer protocols as long as the results
achieved by the simulator are fairly close to reality. 

To simulate large distributed applications and protocols, we have
developed a simulator that scales up to large topologies, large numbers
of connections, and is able to simulate over long periods of simulation
time at varying levels of detail. Narses provides several network
models that trade between fast runtimes and accuracy. The models are
easily interchangeable, so applications can be quickly prototyped
using fast but less accurate models and then simulated with more detailed
network models. Narses thus lies someplace between packet-level simulators
and analytical models in terms of its compromise between speed and
accuracy.

Narses reduces the complexity of simulations by approximating the
behavior of the physical, link, network, and transport layers. Narses
achieves this approximation with two simplifications. First, Narses
reduces memory and computational requirements by simulating at the
flow level rather than at the packet level. Second, Narses assumes
that there are no bottleneck bandwidths in the core of the topology
simulated. This simplification implies that not all topologies are
appropriate for simulation by Narses.

While the coarse-grain study of applications' functional behavior
is a main goal of Narses, we nonetheless present timing results for
our most detailed network model. We have run experiments comparing
its timing results with those of ns. Specifically, we simulate end
hosts transmitting chunks of data back and forth and measure the average
time it takes to complete each flow. Our results show up to 45 times
improvement in simulator run times versus ns while only using 28\%
of the memory consumed by ns. The simulated completion times maintain
an average accuracy of 8\%.

\section{Narses}

Narses is a Java-based network simulator targeted towards large distributed
applications. Narses provides simulated applications with a transport
layer interface through which the application can send and receive
data. The transport layer interface is very similar to a UNIX socket
interface, which allows users to port their simulated applications
easily to a real operating system. 

One goal of Narses is to allow efficient characterization of large
distributed applications by offering several network models that can
be easily interchanged to provide the best tradeoff between speed
and accuracy. For example, Narses offers a {}``naive'' network model
that does not take the effects of cross-traffic into account, but
is very simple and fast to execute. This model can be used to prototype
an application protocol to verify its correctness. When the application
has been debugged, the naive model can be switched with a more detailed
network model to characterize the application's behavior on a more
realistic network. The rest of this paper describes the most detailed
-- and thus most expensive -- of Narses' network models: the bandwidth-share
model.

The main goal of Narses' bandwidth-share model is to provide an approximation
of a TCP-like transport protocol to simulated applications. The model
attempts to replicate the sharing of link bandwidth between independent
flows. TCP flows running over the same link independently interact
with each other and the router queue for that link to divide the bandwidth
of that link amongst themselves. The allocation of bandwidth is not
necessarily fair, but it does allow each flow to use some of the link's
bandwidth. The simulator attempts to reflect this sharing of bandwidth
among flows and to take into account traffic interdependencies.

To achieve this macro-level approximation of TCP, Narses makes two
simplifying assumptions, which are detailed in the next two sections.

\subsection{Flow Simulation}

The first simplifying assumption is that Narses abstracts individual
packet information into the the concept of a flow. In Narses, a flow
is any semantically meaningful chunk of bytes that is passed down
to the transport layer to send to a recipient.

Simulating at the granularity of a flow reduces the runtime of simulations
for several reasons. By grouping packets into flows, Narses reduces
the total number of events in the simulation. Furthermore, the reduction
in the number of events in the system also reduces the memory consumed
by simulations. In packet level simulations, header information for
every packet consumes memory. With flow simulation, the header information
for each packet is subsumed into a more compact, summarized form that
describes the entire flow.

\subsection{Bottleneck Link}

The second simplifying assumption Narses makes is that no bottleneck
link exists in the core of the network throughout the simulation.
This means that transfers between two end hosts are limited by their
first-link connections to the network. For example, a DSL user downloading
content from a CDN would be limited by the speed of the CDN's link
and the DSL link. In this case, the DSL link would most likely be
the limiting factor.

With this assumption, Narses does not have to simulate intermediate
routers in the network. With traditional packet-based simulators,
packets must travel hop by hop through the network and be inspected
by the network layer protocol at each router. Narses only has to decide
how to divide available bandwidth between flows entering and leaving
end hosts.

\subsection{Bandwidth-share Model}

Since Narses uses flows and assumes no intermediate bottleneck links,
its job is simple - allocate bandwidth to all active flows in the
simulation in a way that mimics TCP's bandwidth allocations across
an Internet-like network.

Narses allocates bandwidth to flows using a technique we call \textit{minimum-share
allocation}. Using the bottleneck link assumption, the maximum bandwidth
available to a flow is the minimum of the bandwidth available at its
source and destination. A flow's share of an end host's bandwidth
is the bandwidth of the end host's connection to the network divided
by the number of flows sent or received by that node.

\begin{figure}[htbp]
\includegraphics[  scale=0.9]{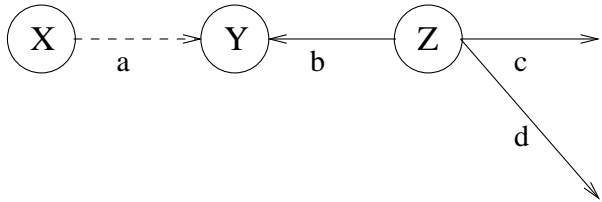}

\caption{\label{fig:reallocation}Communication graph in which flow \textbf{a}
has completed and flow \textbf{b} may now consume more of end host
Y's bandwidth.}
\end{figure}

When a flow starts or completes, the bandwidth shares of every other
flow on the sending node and the receiving node change. Figure~\ref{fig:reallocation}
is a communications graph (not a topology graph) that illustrates
how reallocation is done. %
 In this description, we refer to an end host's bandwidth as the bandwidth
of an end host's first link connection to the network. In the figure,
end host \textbf{Y} receives two flows \textbf{a} and \textbf{b}.
Thus the bandwidth share at \textbf{Y} for all of its flows is \textbf{Y}'s
bandwidth divided by 2. End host \textbf{Z} sends three flows, so
the bandwidth share for each flow at \textbf{Z} is the bandwidth of
\textbf{Z} divided by 3. Therefore, flow \textbf{b}'s bandwidth is
calculated using the minimum share allocation: $\min (\frac{_{Z_{bandwidth}}}{3},\frac{_{Y_{bandwidth}}}{2})$.

Assume now that flow \textbf{a} completes. Only one flow is received
by \textbf{Y}, so \textbf{b}'s bandwidth share at \textbf{Y} increases.
\textbf{b}'s bandwidth allocation is now $\min (\frac{_{Z_{bandwidth}}}{3},\frac{_{Y_{bandwidth}}}{1})$.
Note that since the number of flows entering or leaving \textbf{Z}
did not change, flows \textbf{c} and \textbf{d} do not need to be
reallocated bandwidth. One benefit of the bandwidth-share model is
that the initiation or completion of a flow causes only a limited
number of flows to be reallocated, as seen in the above example.

\subsection{Limitations}

Because of the simplifying assumptions we make, Narses has a number
of limitations that do not apply to general purpose packet-based network
simulators.

First, Narses cannot simulate topologies with bottleneck links that
are not first-hop links. This means that it is restricted to Internet-like
hierarchical topologies without internal bottlenecks.

Second, Narses cannot simulate different bit error rates for physical
channels. This is not too much of a limitation for wired networks
where bit error rates are typically very low and unchanging. However,
wireless networks have much higher bit error rates and bursty losses,
all of which wreak havoc with traditional TCP implementations. Because
of these factors, we do not use Narses to simulate wireless networks.

Finally, Narses is targeted towards applications, so simulations concerning
lower-layer protocol dynamics are not possible with Narses. For lower
layer information, traditional network simulators should be used instead.

\section{Evaluation}

In this section we explain our experimental methodology and give preliminary
results using the bandwidth-share model.

\subsection{Methodology}

We evaluate the accuracy of Narses' bandwidth-share model by running
identical simulation scenarios in Narses and in ns and comparing the
results. Each simulation is generated by transferring flows of data
between random end hosts. In the ns simulations, a flow is simulated
by sending a chunk of bytes over a TCP connection. The flows are all
of the same size.

In each simulation, we measure the simulated completion time of each
flow. By simulated completion time of a flow, we mean the latency
in simulation time units from start to finish of the flow. All simulated
completion times were averaged together and compared with the average
simulated completion time of the other simulator. We measure simulated
flow completion times because applications regard each flow as a single
message that is to be serviced when it has been completely received.

The topology was created using the GT-ITM topology generator\cite{Zegu1996}.
The topology is a transit-stub model network containing 600 nodes
with no bottleneck links between end hosts. Throughout the simulations,
the average round-trip path latency in ns is 96ms with a maximum latency
of 190 ms. The average round-trip path latency in Narses is 88ms with
a maximum latency of 158ms. This discrepancy is due to the fact that
the ns simulations use hierarchical addressing, which lowers the routing
table size per node from $O(n)$ to $O(\log (n))$ but sometimes chooses
non-optimal routes. Narses uses an optimal minimum spanning tree rooted
at each node which consumes $O(n)$ space per node. The discrepancy
in round-trip times adds some error into Narses' accuracy results
but does not affect the simulation runtime results or memory consumption
results.

Our simulation machine is a dual 2.4GHz Xeon with 2GB of RAM running
RedHat Linux 8.0.

\subsection{Accuracy\label{sub:Accuracy}}

\begin{figure}[htbp]
\includegraphics[  scale=0.62]{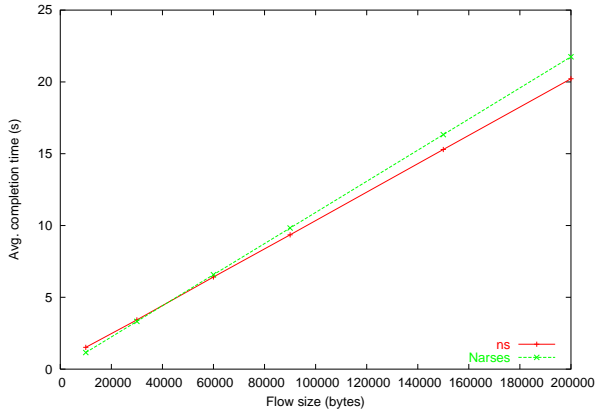}

\caption{\label{fig:acc10000}Average flow completion times (in simulation
time). Flow sizes vary between 10KB and 200KB.}
\end{figure}
 To test the accuracy of Narses, we simulated 10000 simultaneous flows
of the same size and measured their simulated completion times. We
then varied the size of the flows and calculated the average simulated
completion time for each flow size. Figure~\ref{fig:acc10000} shows
these results. For a flow size of 200KB, for example, Narses' results
differ by 1.53s, which is a 7.6\% difference.

\subsection{Runtime}

\begin{figure}[htbp]
\includegraphics[  scale=0.62]{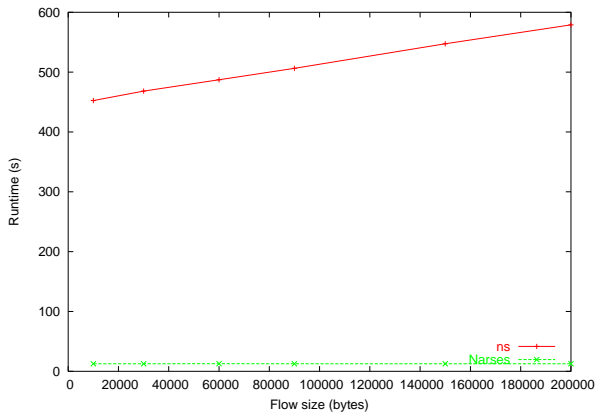}

\caption{\label{fig:run10000}Runtime of the simulations.}
\end{figure}

Figure~\ref{fig:run10000} shows the runtime of the same experiment
performed in section \ref{sub:Accuracy}. As flow size increases,
ns slows down because there are more total packets that must be dealt
with in the course of the simulation. As described earlier, the larger
number of packets increases the main event queue size. Additionally,
ns must perform routing and TCP computations on a larger number of
packets. Narses' runtime remains constant and is 45 times faster than
ns with a flow size of 200KB. %
 It should be noted that Narses is implemented in Java while ns is
implemented in C++, which gives ns a runtime advantage.

\subsection{Memory Consumption}

\begin{figure}[htbp]
\includegraphics[  scale=0.62]{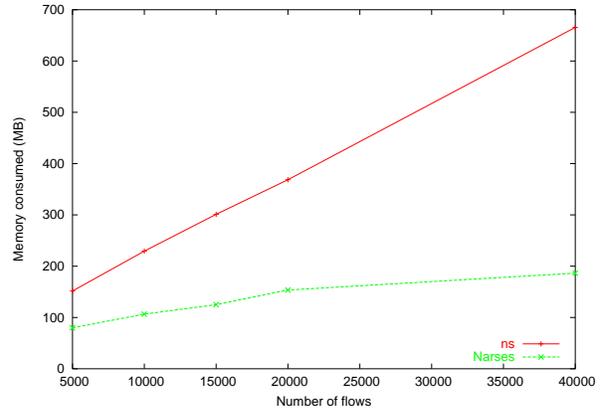}

\caption{\label{fig:mem10000}Memory consumption.}
\end{figure}

Figure~\ref{fig:mem10000} shows the memory consumed as the flow
size is held constant at 200KB and the number of flows is varied from
5000 random flows to 40000 random flows. At 40000 flows, Narses consumes
28\% of the memory that ns consumes. %

\section{Related Work}

The most closely related work to this work is the FlowSim simulator
by Ahn and Danzig\cite{Ahn1996}. FlowSim is a packet-level discrete-event
simulator that aggregates closely spaced packets on a link with the
same source and destination into a packet train. Packets that have
the same source and destination but are not closely spaced are not
aggregated, whereas Narses aggregates all packet information, regardless
of packet spacing.

Another interesting abstraction technique was investigated by Huang
and Heidemann \cite{huang2001}. They develop a technique to simulate
background TCP traffic in a memory-efficient manner. To save memory,
they model TCP connections using finite-state machines that approximate
the behavior of TCP. The FSM representation of TCP means that TCP
state variables, timers, and other space is not needed for most connections.

\section{Future Work}

Since Narses is written in Java, the simulator relies on Java's garbage
collector to manage memory. In simulations with large numbers of flows,
many transient flow objects are created and left to the garbage collector
to reclaim. We are currently implementing a memory manager that reuses
frequently used objects. We expect this to improve runtime performance
and possibly decrease the amount of memory consumed.

The real utility of Narses is in measuring large distributed applications.
Narses is currently being used to study several distributed systems,
including CUP \cite{rous2002}. We also plan to implement several
large distributed applications such as Gnutella that have been well
studied. We will simulate these applications using Narses to check
whether Narses reproduces behaviors known to exist in the real applications.

Another desired direction for Narses is to explore network emulation.
Other discrete event simulators such as ns have support for emulating
the behavior of a real network in real time. We wish to investigate
the feasibility of network emulation using a flow-based simulator.

\section{Conclusion}

We have developed an application-level simulator, Narses, that can
simulate large numbers of hosts, large numbers of concurrent connections,
and long periods of time. We have built a discrete event, TCP-flow
based simulator that elides individual packet information but still
reflects the impact of traffic interdependences. Our results show
our technique to be reasonably accurate while being more scalable
than packet-level simulation.

\section{Acknowledgements}

This project has been funded in part by MURI award number F49620-00-1-0330.


\begin{thebibliography}{1}
\bibitem{Ahn1996}J. Ahn and P.B. Danzig. Speedup vs. Simulation Granularity. \textit{IEEE/ACM
Transactions on Networking}. 4(5):743-757, October 1996.
\bibitem{clip2}DSS Group. Gnutella: To the Bandwidth Barrier and Beyond. Nov. 6,
2000.
\bibitem{hong2001}Theodore Hong. Chapter 14: Performance. \textit{Peer-to-Peer: Harnessing
the Power of Disruptive Technologies}, edited by Andy Oram. O'Reilly
and Associates: Sebastopol, CA (2001).
\bibitem{huang2001}Polly Huang and John Heidemann. Capturing TCP Burstiness in Light-weight
Simulations. \textit{Proceedings of the SCS Conference on Communication
Networks and Distributed Systems Modeling and Simulation}, pp. 90-96.
Phoenix, Arizona, USA. January, 2001.
\bibitem{ns}Ns Project http://www.isi.edu/nsnam/ns/
\bibitem{rous2002}Mema Roussopoulos and Mary Baker, CUP: Controlled Update Propagation
in Peer-to-Peer Networks. Technical Report cs.NI/0202008, http://arXiv.org/abs/cs.NI/0202008,
February, 2002.
\bibitem{Zegu1996}Ellen W. Zegura, Ken Calvert and S. Bhattacharjee. How to Model an
Internetwork. \textit{Proceedings of IEEE Infocom} '96, San Francisco,
CA. 
\bibitem{zhan2002}H. Zhang, A. Goel, R. Govindan. Using the Small World Model to Improve
Freenet Performance. \textit{Proceedings of IEEE Infocom}, June 2002.\end{thebibliography}
\end{document}